\documentclass[runningheads]{llncs}
\usepackage[T1]{fontenc}
\usepackage{graphicx}
\usepackage{orcidlink}
\usepackage{amsmath}
\usepackage{subcaption}
\usepackage[nolist]{acronym}
\usepackage{cleveref}
\usepackage{listings}
\usepackage{xcolor}
\usepackage{float}
\usepackage{tabularx}
\usepackage{booktabs}

\newcolumntype{Y}{>{\centering\arraybackslash}X}

\begin{acronym}
\acro{llm}[LLM]{large language model}
\acro{amt}[AMT]{asynchronous many-task}
\acro{blas}[BLAS]{Basic Linear Algebra Subprograms}
\acro{ttg}[TTG]{Template Task Graph}
\acro{rss}[RSS]{resident set size}
\end{acronym}
\begin{document}

\title{Can Agentic LLMs Write Task-Based Parallel Code? A Study of OpenMP Tasking, HPX, StarPU and TTG}

\titlerunning{Can Agentic LLMs Write Task-Based Parallel Code?}
\author{Linus Bantel\inst{1}\orcidlink{0009-0009-0247-4748} \and
Anna-Lena Roth\inst{2}\orcidlink{0000-0002-6463-3486} \and
Tim Th\"uring\inst{1}\orcidlink{0009-0006-1762-8429} \and
Jonas Posner\inst{2}\orcidlink{0000-0002-6491-1626} \and
Dirk Pfl\"uger\inst{1}\orcidlink{0000-0002-4360-0212}}
\authorrunning{Bantel et al.}
\institute{Institute for Parallel and Distributed Systems, University of Stuttgart, Stuttgart, Germany\\
\email{\{linus.bantel, tim.thuering, dirk.pflueger\}@ipvs.uni-stuttgart.de}
\and
Fulda University of Applied Sciences, Fulda, Germany\\
\email{\{anna-lena.roth, jonas.posner\}@informatik.hs-fulda.de}
}
\maketitle
\begin{abstract}
Task-based programming offers more expressive power than loop-level parallelism, but the runtimes providing it differ widely in API style, maturity, and the amount of public code to learn from.
This paper asks how well \ac{llm} agents write C\texttt{++} for four of them: OpenMP Tasking, HPX, StarPU, and \ac{ttg}.
Instead of single one-shot prompts, we use an agentic workflow in which the model edits files, compiles, and runs its kernels until it reproduces a reference output.
Five locally hosted open-weight \acp{llm} implement seven HPC kernels, including matrix multiplication, Cholesky and LU factorization, in all four frameworks and a serial baseline to verify their understanding of the problem in general.
Every kernel is evaluated independently for correctness at each concurrency level, strong scaling, and peak memory.
Model capability dominates the outcome: three models produce correct, genuinely task-parallel kernels in 57--82\% of the cases, while two models rarely compile their own code and almost never succeed.
For the capable models, OpenMP Tasking and HPX are handled most reliably, StarPU is close behind, and \ac{ttg} trails.
The best generated GEMM and factorization kernels match or exceed the multithreaded OpenBLAS/LAPACK reference. However, correctness overstates parallel quality: several ``parallel'' kernels contain no parallelism at all; and serialized task graphs and unsuitable task granularity remain common.
\end{abstract}
\section{Introduction}
\acp{llm} have transformed software engineering, but their ability to produce correct and performant parallel programs remains limited~\cite{Nichols_2024}.
In previous work, we evaluated \ac{llm}-generated task-based parallel code from single prompts for OpenMP Tasking, C\texttt{++} standard parallelism, and HPX~\cite{bantel_prompts}.
Many programs required human fixes before they even compiled, and HPX in particular suffered from outdated or hallucinated APIs.

Since then, \acp{llm} have shifted from single prompts to agents that read files, edit code, and run a compiler or the program itself~\cite{react,sweagent}.
This closes the loop we had to close manually and should mitigate errors such as outdated APIs.
Whether it also resolves semantic failures, e.g., wrong dependencies, deadlocks, or poor scaling, remains unclear.

This is particularly relevant for \ac{amt} runtimes, which enable efficient irregular parallelism in large-scale applications, e.g., HPX on GPUs for stellar mergers~\cite{daiss2026irregular}.
However, they are harder to program than fork-join models: the programmer must express an explicit dependency graph, and for some runtimes little public code exists to learn from.
The only evidence so far comes from Julia, where agents produce executable code with the task-based Dagger.jl for small inputs but often fail at scale due to deadlocks, oversubscription, and weak task dependencies~\cite{bantel_julia}.
It is unknown whether the Dagger.jl failures are framework-specific, language specific, or typical for task-based programming.

We therefore study \ac{llm} agents writing task-parallel C\texttt{++} for OpenMP Tasking, HPX, StarPU, and \ac{ttg}, which have very different programming models.
We use a plain agentic workflow without custom tools, five locally hosted open-weight \acp{llm}, and seven benchmarks ranging from an embarrassingly parallel $\pi$ approximation to tiled LU factorization.
Generated kernels are evaluated independently of the agent for correctness at every concurrency level, strong scaling, and peak memory, and we analyze code and agent sessions to explain failures.
Specifically, we answer:
\begin{itemize}
    \item \textbf{RQ1:} Do agents produce correct kernels for each framework?
    \item \textbf{RQ2:} Do the correct kernels scale, and how much memory do they need?
    \item \textbf{RQ3:} Why do kernels fail or not scale, and do the causes differ between frameworks?
\end{itemize}

\section{Related Work}\label{sec:related_work}

\acp{llm} such as Qwen~\cite{qwen3} generate correct code across many programming languages~\cite{code_gen_1,code_gen_3}, but HPC code remains a challenge.
Early evaluations of OpenMP, OpenACC, CUDA, and MPI kernels found that success depends on the availability of training data~\cite{parallel_kernels,HPC-Coder,parallel_kernel_3}, and ParEval~\cite{Nichols_2024} showed that models perform substantially worse on parallel than on serial problems and often produce correct but inefficient code.
Our previous study extended these evaluations to task-based frameworks, including HPX and OpenMP Tasking, and to strong and weak scaling~\cite{bantel_prompts}.
In software engineering, the focus has moved to agents that interleave reasoning with tool calls~\cite{react} and resolve real repository issues~\cite{swebench,sweagent}, evaluated by functional tests but not by parallel performance.
Agentic generation of parallel Julia code has been evaluated on a supercomputer~\cite{bantel_julia}, but no work compares \acp{llm} as coding agents across task-based C\texttt{++} runtimes such as HPX~\cite{hpx}, StarPU~\cite{starpu}, or \ac{ttg}~\cite{ttg}.

\section{Methodology}\label{sec:methods}

Our benchmark problems vary in algorithmic structure, dependency complexity, sensitivity to compute, and memory bandwidth limits. The factorizations follow tiled dense linear algebra~\cite{tiled_la}.
The problem sizes used for the evaluation are given in parentheses; for the dense linear algebra problems, they are chosen as powers of two such that a single-threaded OpenBLAS/LAPACK reference takes several seconds.
OpenBLAS and LAPACK also serve as a reference to compare the generated kernels to.
\begin{enumerate}
    \item \textbf{Approximation of $\pi$} by midpoint integration of $4/(1+x^2)$ over $[0,1]$, representing independent tasks with a single reduction ($2^{32}$ steps).
    \item \textbf{AXPY} and \textbf{GEMV}, memory-bandwidth-bound vector and matrix-vector operations with independent output blocks ($n = 2^{30}$ and $2^{15} \times 2^{15}$).
    \item \textbf{Matrix-matrix multiplication (GEMM)}, a compute-bound kernel with high arithmetic intensity and independent output tiles ($N = 2^{12}$).
    \item \textbf{Cholesky} and \textbf{LU factorization} (without pivoting), which require a task graph with a sequential critical path and four distinct task types in the tiled formulation~\cite{tiled_la} ($N = 2^{13}$).
    \item \textbf{Linear solve}, i.e., a factorization followed by forward and backward substitution, combining two dependent phases ($N = 2^{13}$).
\end{enumerate}

The agent implements each kernel as a serial program, as a baseline to verify the models understand the problem, and in each of the following four frameworks:
\begin{enumerate}
    \item \textbf{OpenMP Tasking}~\cite{openmp_tasks} is a directive-based extension of the most widely used shared-memory model. Dependencies are expressed with \texttt{depend} clauses on memory addresses.
    \item \textbf{HPX}~\cite{hpx} is an \ac{amt} runtime whose API mirrors the C\texttt{++} standard library. Tasks are futures, and dependencies are expressed through \texttt{dataflow} and continuations.
    \item \textbf{StarPU}~\cite{starpu} is a C runtime in which data is registered as handles and tasks are instances of codelets. Dependencies are inferred from the access modes of the handles a task uses.
    \item \textbf{\ac{ttg}}~\cite{ttg} is the youngest of the four frameworks. Programs are expressed as a template task graph of operations connected by typed edges, and are executed by the PaRSEC runtime~\cite{parsec}.
\end{enumerate}

Together, they range from directives over futures to data handles and flow graphs, and from abundant public example code for OpenMP to scarce examples for \ac{ttg}.

\subsection{Agentic Workflow}

We use the open-source coding agent opencode~\cite{opencode} (version 1.18.31) with only its standard tools: reading, writing, and editing files, executing shell commands, fetching web pages, and maintaining a todo list.
We deliberately provide no custom tools, e.g., for documentation search or verification, to evaluate a workflow that any developer can reproduce.
The agent runs inside an Apptainer container with GCC, CMake, OpenBLAS, LAPACK, and the four frameworks, and has no access to the evaluation data.

All variants share a fixed contract.
The agent writes a single file \texttt{kernel.cpp} defining \texttt{read\_input}, \texttt{run\_kernel}, and \texttt{write\_output}.
A fixed harness provides \texttt{main}, initializes and finalizes the runtime, sizes the worker pools to the requested number of threads, and measures only the runtime of \texttt{run\_kernel}.
The agent receives the build recipe, the harness, the data generator, a small sample input with the expected output, and, for the parallel variants, the serial reference kernel.
\ac{blas} and LAPACK routines may be used as block-local building blocks, but not as a single call that solves the whole problem.

Besides the problem-specific task prompt, a system prompt summarizes principles of task-based HPC programming, e.g., expressing dependencies instead of barriers, choosing a task granularity, and avoiding data races in reductions, and gives short idiomatic notes per framework.
It also instructs the agent to design the task graph first, then build, run, and verify its kernel against the sample output for several thread counts.
Sessions are limited to two hours, after which the last \texttt{kernel.cpp} is kept.
We generate one kernel per model, problem, and variant, i.e., 35 per model and 175 in total.

\subsection{Models}

We evaluate five open-weight \acp{llm} that are served locally with llama.cpp~\cite{llamacpp} on a node with four NVIDIA V100 GPUs with 32\,GB each.
We refer to them by their server aliases: qwen3.8:27b (Qwen3.8-27B, Q4\_K\_M quantization), ornith:35b (Ornith-1.5-35B, Q4\_K\_M), nemotron3:30b (Nemotron-3-Nano-30B-A3B, IQ4\_XS), nemotron-3-nano:4b (NVIDIA-Nemotron-3-Nano-4B, Q8\_0), and qwen3.6:35b.
All models use a context window of 262,144 tokens.

\subsection{Evaluation}\label{sec:metrics}

The evaluation is fully separated from the generation and uses new inputs of the size given in \Cref{sec:methods}, whose reference output the agent has never seen.
Each kernel is run for $p \in \{1,2,4,8,16,32,64\}$ threads, and we report the geometric mean of two timed runs of \texttt{run\_kernel}, aborting runs after 900\,s.
A kernel is correct if its output matches the reference (relative tolerance $10^{-5}$, absolute $10^{-8}$) for every $p$.
We apply no fixes and further classify correct parallel kernels without any task parallelism as \emph{correct, no parallelism}; contract violations such as hard-coded results count as wrong.
For the PCGQS~\cite{bantel_prompts}, we use binary correctness and the strong scaling efficiency $S_{\text{strong}}$ averaged over all $p$, scoring kernels without task parallelism as incorrect.

\section{Results}\label{sec:results}

All kernels are evaluated on a system equipped with two AMD EPYC 7542 CPUs with 32 cores each and 1\,TB of main memory.
We use the same container as for the generation, with GCC 13.3.0, HPX 1.10.0, StarPU 1.4.7, \ac{ttg} 1.0.0 with its PaRSEC backend, and OpenBLAS 0.3.26, which the harness restricts to a single thread.
As a reference for the dense linear algebra problems, we implement each of them with a single multithreaded OpenBLAS or LAPACK call within the same harness.

\subsection{Correctness}

\begin{figure}[t]
    \centering
    \begin{subfigure}[t]{0.48\textwidth}
        \centering
        \includegraphics[width=0.99\linewidth]{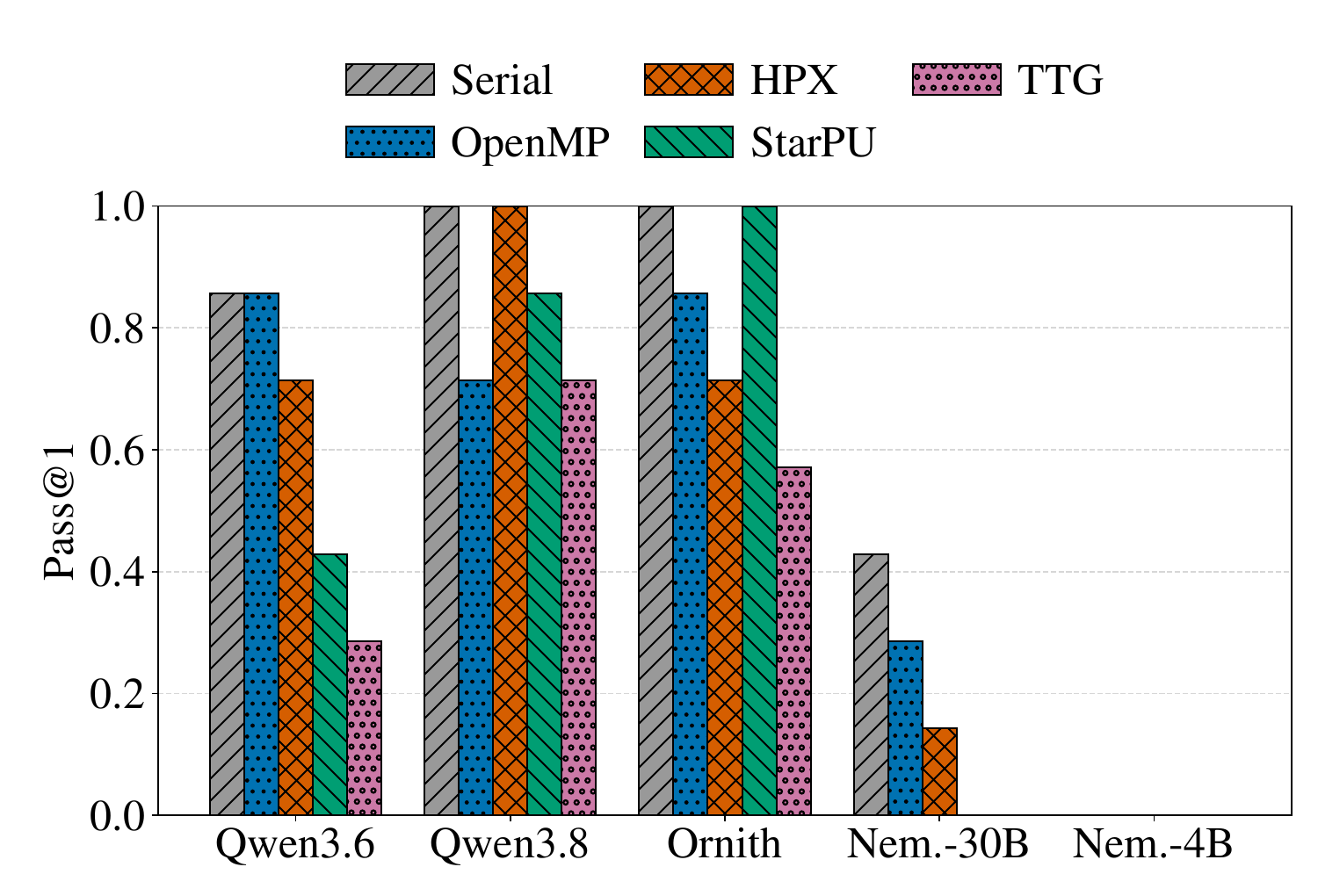}
        \caption{Grouped by framework.}
        \label{fig:pass_at_1_framework}
    \end{subfigure}
    \begin{subfigure}[t]{0.48\textwidth}
        \centering
        \includegraphics[width=0.99\linewidth]{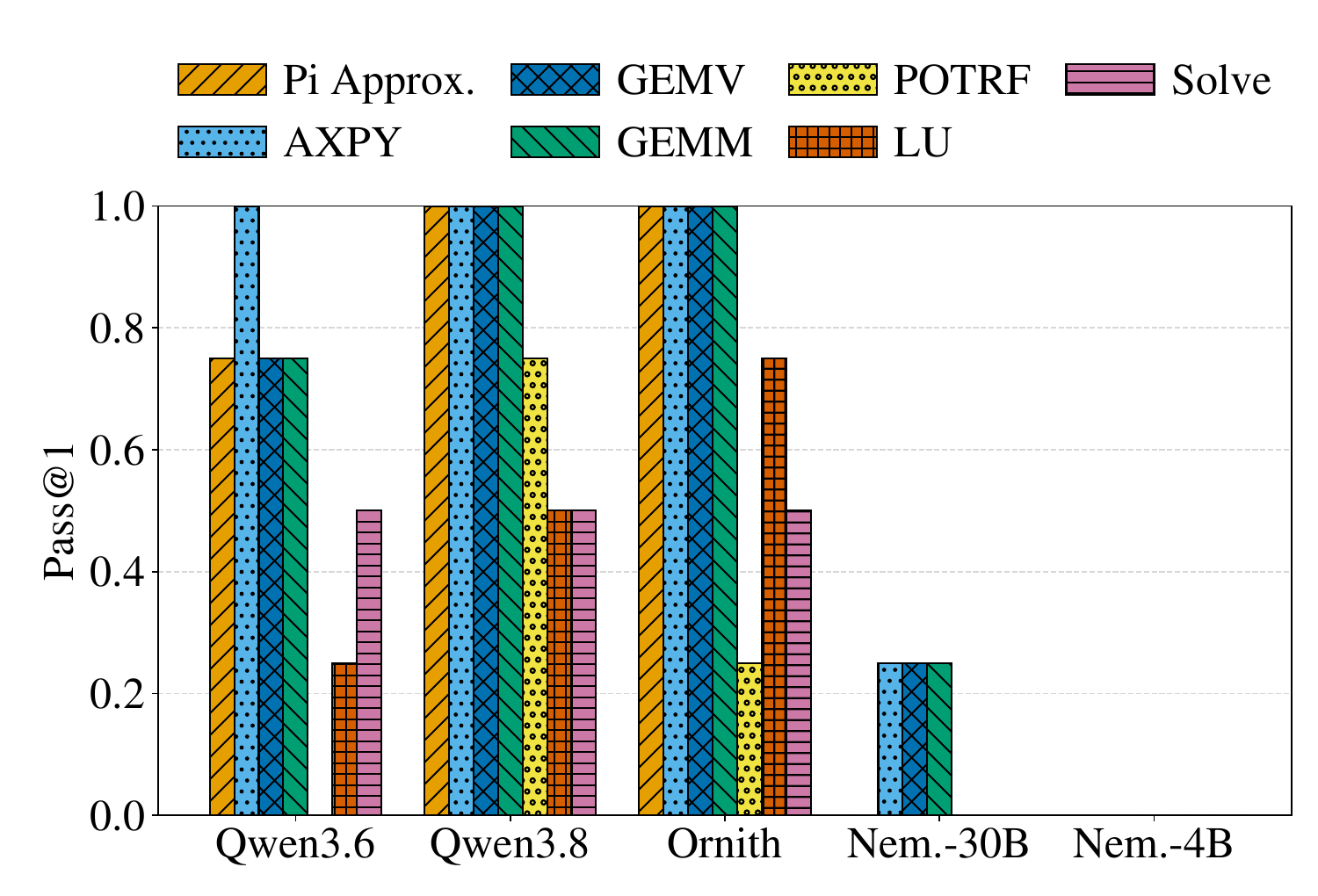}
        \caption{Grouped by problem (parallel variants).}
        \label{fig:pass_at_1_problem}
    \end{subfigure}
    \begin{subfigure}[t]{0.48\textwidth}
        \centering
        \includegraphics[width=0.99\linewidth]{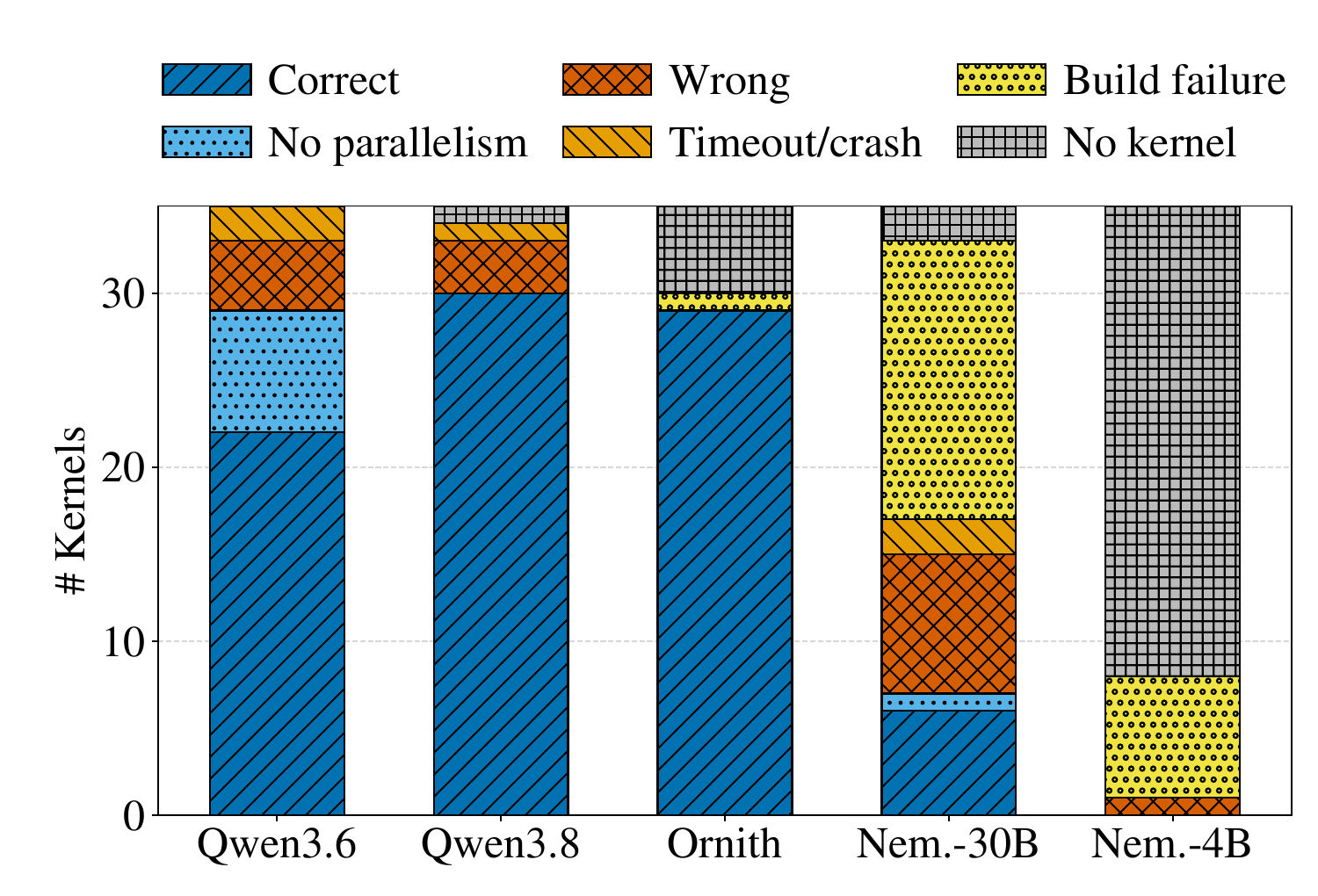}
        \caption{Outcome of all 35 kernels.}
        \label{fig:outcomes_model}
    \end{subfigure}
    \begin{subfigure}[t]{0.48\textwidth}
        \centering
        \includegraphics[width=0.99\linewidth]{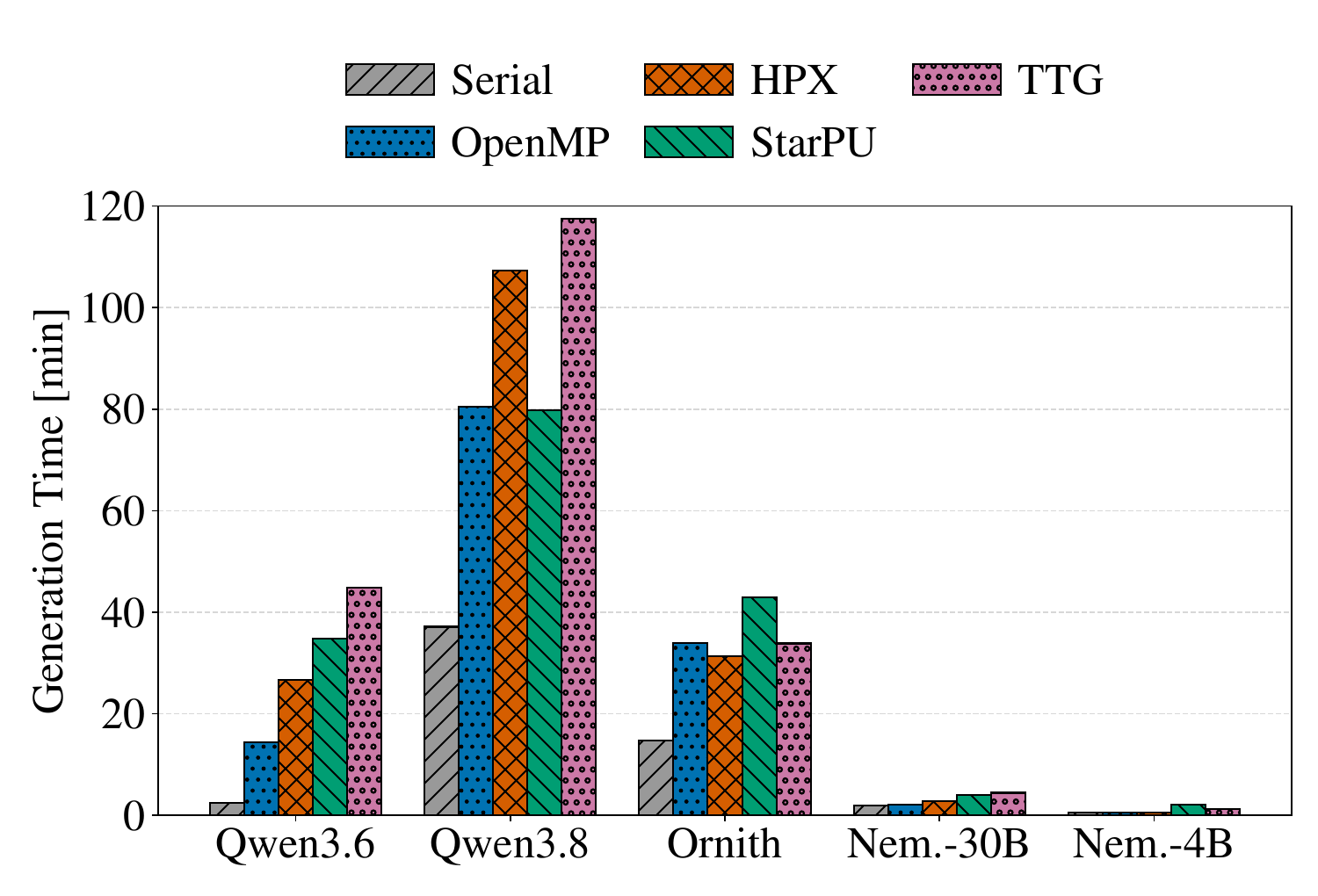}
        \caption{Mean duration of the agent sessions.}
        \label{fig:gen_time}
    \end{subfigure}
    \caption{Pass@1 metric for the respective \acp{llm} without any human correction (a, b), outcome of all generated kernels (c), and duration of the agent sessions (d). Kernels without task parallelism count as incorrect.}
    \label{fig:outcomes}
\end{figure}

\Cref{fig:outcomes} shows the pass@1 metric and the outcome of all 175 generated kernels.
The differences between the models dominate all other factors.
qwen3.8:27b and ornith:35b produce 30 and 29 correct kernels out of 35, respectively, and all of them contain genuine task parallelism.
qwen3.6:35b produces 29 kernels that pass the correctness check, but seven of its parallel kernels contain no parallelism.
For example, its HPX and \ac{ttg} variants of LU factorization are plain serial loops, and its StarPU variants of Cholesky and LU factorization submit the entire serial algorithm as a single task.
In stark contrast, nemotron3:30b produces only six correct kernels, and nemotron-3-nano:4b produces none.
The only kernel of nemotron-3-nano:4b that passes the check returns the constant $3.14159\ldots$, and two kernels of nemotron3:30b call a single forbidden LAPACK routine.

As shown in \Cref{fig:pass_at_1_framework}, the three strongest models produce correct serial baselines for all problems, except for the naive serial GEMM of qwen3.6:35b, which exceeds the time limit.
Aggregated over these models, OpenMP Tasking and HPX each yield 17 correct parallel kernels out of 21, followed by StarPU with 16 and \ac{ttg} with only 11.
The ranking differs between the models; however, qwen3.8:27b solves all problems with HPX, ornith:35b solves all problems with StarPU, and qwen3.6:35b performs best with OpenMP.
Apart from naive serial GEMM kernels, \ac{ttg} is also the only framework with timeouts and crashes and the most frequent fallback to serial code.
Thus, the ranking reflects the maturity and public visibility of the frameworks, but the differences among OpenMP, HPX, and StarPU are small, in contrast to our previous study, in which HPX code from single prompts often failed to compile~\cite{bantel_prompts}.

\Cref{fig:pass_at_1_problem} reveals a clear dependence on the problem.
The three strongest models reliably solve AXPY (12 of 12 parallel variants), GEMV (11), and GEMM (11), whereas only four of twelve parallel kernels are correct and genuinely parallel for Cholesky factorization and six each for LU factorization and the linear solve.

\begin{table}[t]
\centering
\caption{Root causes of the 42 kernels that are wrong, time out, or do not build, determined by manual inspection and rebuilding.}
\label{tab:errors}
\begin{tabularx}{\textwidth}{Xr}
\toprule
\textbf{Root cause} & \textbf{\#Kernels} \\
\midrule
\textit{Build failures} & 24 \\
\quad Non-existent API, e.g., headers \texttt{hpx/parallel.hpp}, \texttt{ttg/graph.hpp} & 10 \\
\quad General C\texttt{++} errors, e.g., declaration order, mismatched types & 7 \\
\quad Wrong use of an existing API, e.g., template arguments of \texttt{ttg::make\_tt} & 4 \\
\quad Incomplete code & 3 \\
\midrule
\textit{Wrong results and timeouts} & 18 \\
\quad Algorithm or indexing errors, e.g., integer truncation, wrong tile offsets & 7 \\
\quad Wrong output format, e.g., a 32-bit header instead of 64-bit & 4 \\
\quad Missing or incorrect task dependencies & 2 \\
\quad Deadlock or heap corruption leading to a timeout & 2 \\
\quad Contract violations, e.g., hard-coded result & 3 \\
\bottomrule
\end{tabularx}
\end{table}

To understand the failures, we rebuilt every kernel that did not build and inspected the code of every kernel that produced wrong results or timed out.
\Cref{tab:errors} summarizes the root causes.
Almost all build failures stem from the two weakest models, and ten of them are caused by hallucinated APIs, mostly headers and functions of HPX and StarPU that do not exist.
Only two of the 24 kernels that do not build were ever compiled by the agent during its session.
Among the wrong results, algorithm and indexing errors dominate.
For example, the StarPU variant of the $\pi$ approximation by qwen3.6:35b limits the number of tasks with \texttt{std::min(num\_threads, (int)num\_steps)}.
For $2^{32}$ steps, the cast yields 0, and the subsequent guard \texttt{if (num\_partial < 1) num\_partial = 1;} reduces the kernel to a single task that integrates only the first of $p$ blocks, which is correct only for $p = 1$.
Missing task dependencies, i.e., the error class that is specific to task-based programming, are rare but subtle.
For example, the OpenMP Cholesky kernel by qwen3.8:27b names local pointer variables in its \texttt{depend} clauses instead of the tiles they point to, so the runtime tracks reused stack addresses and all ordering between tasks is lost.
Both deadlock and heap corruption occur in \ac{ttg} kernels: one waits on a task that is never created and times out, and the other writes through a wrong tile offset, which corrupts the heap and yields a wrong result.
In addition, three GEMM kernels of qwen3.6:35b and nemotron3:30b with naive loops exceed the time limit of 900\,s, and one \ac{ttg} Cholesky kernel crashes at the large problem size.

\subsection{Agent Behavior}

The large gap between the models is explained by how they use the agentic loop.
The three strongest models read the provided files, compile their kernel several times, and test it against the sample output, with a median of 66 to 92 tool calls and four to seven build commands per session.
The two Nemotron models, in contrast, almost never compile their code.
nemotron3:30b ends 29 of its 35 sessions without a single build command, and nemotron-3-nano:4b typically ends its session after reading one file, without writing any code.
Across all models, 90 of the 108 kernels that the agent built at least once pass the correctness check, compared with 5 of the 67 kernels that were never built.
Thus, the agentic loop is highly effective, but only if the model actually uses it.

More effort, however, does not translate into better results, as shown in \Cref{fig:gen_time}.
For both Qwen models, the \ac{ttg} sessions take longest.
qwen3.8:27b spends 49 hours on its 35 kernels and is stopped at the two-hour limit in 17 sessions, often after a correct kernel already existed.
Its sessions show extensive micro-optimization, e.g., inspecting assembly code of the serial $\pi$ approximation.
qwen3.6:35b, which needs only 14 hours, reaches the same number of passing kernels.
Even the strongest models occasionally end a session without writing any code, i.e., five sessions of ornith:35b and one of qwen3.8:27b, in some cases at the output limit of the model.

\subsection{Scaling}

\begin{figure}[!t]
    \centering
    \includegraphics[width=\textwidth]{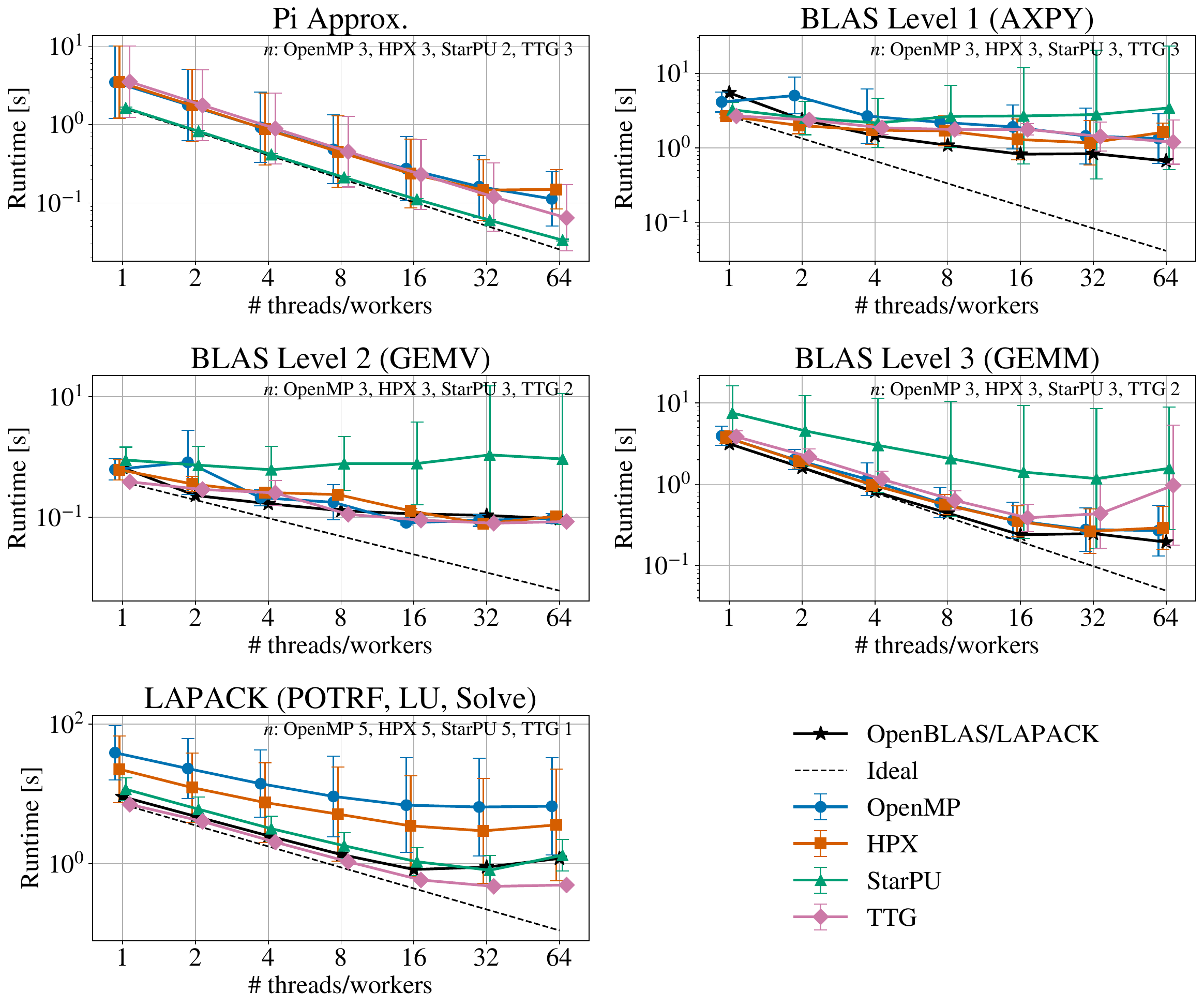}
    \caption{Runtime of the correct and genuinely parallel kernels of the three strongest models (geometric mean).}
    \label{fig:runtime}
\end{figure}

\Cref{fig:runtime} shows the absolute runtimes per problem class.
As expected, the approximation of $\pi$ scales best, with speedups between 48 and 61 on 64 cores for 7 of the 11 correct kernels.
The memory-bound AXPY and GEMV kernels reach median speedups of only 1.4 and up to 8.5, respectively, limited by the memory bandwidth.
For the compute-bound GEMM and the factorizations, the median speedup of the 28 correct and genuinely parallel kernels on 64 cores is 10.2, and most of them peak at 32 cores, similar to the OpenBLAS and LAPACK references.
In terms of $S_{\text{strong}}$~\cite{bantel_prompts}, no framework is consistently best: OpenMP yields the highest efficiency for qwen3.6:35b, HPX for qwen3.8:27b, and StarPU for ornith:35b, while \ac{ttg} is among the lowest for all three models.

Note that $p$ only sets the size of the worker pool of each framework, i.e., the number of HPX worker threads, StarPU CPU workers, PaRSEC threads for \ac{ttg}, or the size of the OpenMP team, while the number of tasks follows from the decomposition chosen by the kernel itself.
Some kernels keep the tile size fixed and submit the same task graph for every $p$, so that only the number of workers executing it changes.
Others derive the block size from $p$ to create a multiple of $p$ tasks, so that the task granularity changes along the scaling curve, which is beneficial as long as the block size is bounded from below.

The GEMM and factorization kernels that scale best share three properties: they decompose the matrix into two-dimensional tiles that remain large enough for efficient \ac{blas} calls at all $p$, they compute each tile with a single level-3 \ac{blas} call, and they express dependencies per tile, i.e., via \texttt{dataflow}, StarPU data handles, or \ac{ttg} edges, so that the next panel starts while the trailing updates of the previous one are still running.
The poorly scaling kernels violate at least one of these properties.
First, some kernels serialize their tasks through the framework itself: the StarPU variants of AXPY and GEMV by qwen3.6:35b register the whole output vector as a single data handle that every task writes, which StarPU's sequential consistency serializes.
Second, the task granularity is unsuitable.
Several kernels derive their block size from the number of threads and shrink it as $p$ grows, e.g., the \ac{ttg} variant of GEMM by qwen3.8:27b ends up with millions of tiny tasks, while others compute each output tile with many small \ac{blas} calls that re-read the input tiles and are limited by the memory bandwidth instead of the floating-point throughput.
Third, the factorizations by qwen3.6:35b use the unblocked algorithm, i.e., a rank-1 update of the whole trailing matrix per column followed by a \texttt{taskwait} or \texttt{when\_all().get()}, so that each of the $N$ steps streams the matrix through memory and synchronizes all workers, and even the tiled HPX solve by qwen3.8:27b factorizes each panel column by column with a blocking join, which places a memory-bound phase on the critical path.
Fourth, four AXPY kernels zero-initialize the output vector serially inside \texttt{run\_kernel}, whereas the kernels that allocate uninitialized memory let each task touch its part first.
Finally, slow serial code scales better: the non-vectorized $\pi$ kernels by qwen3.6:35b need ten times longer per step and therefore scale almost ideally, whereas the vectorized kernels of the other models finish within tens of milliseconds on 64 cores, where the fixed costs of the runtimes become visible, so we also report absolute runtimes.

\begin{figure}[p]
    \centering
    \includegraphics[width=.95\textwidth]{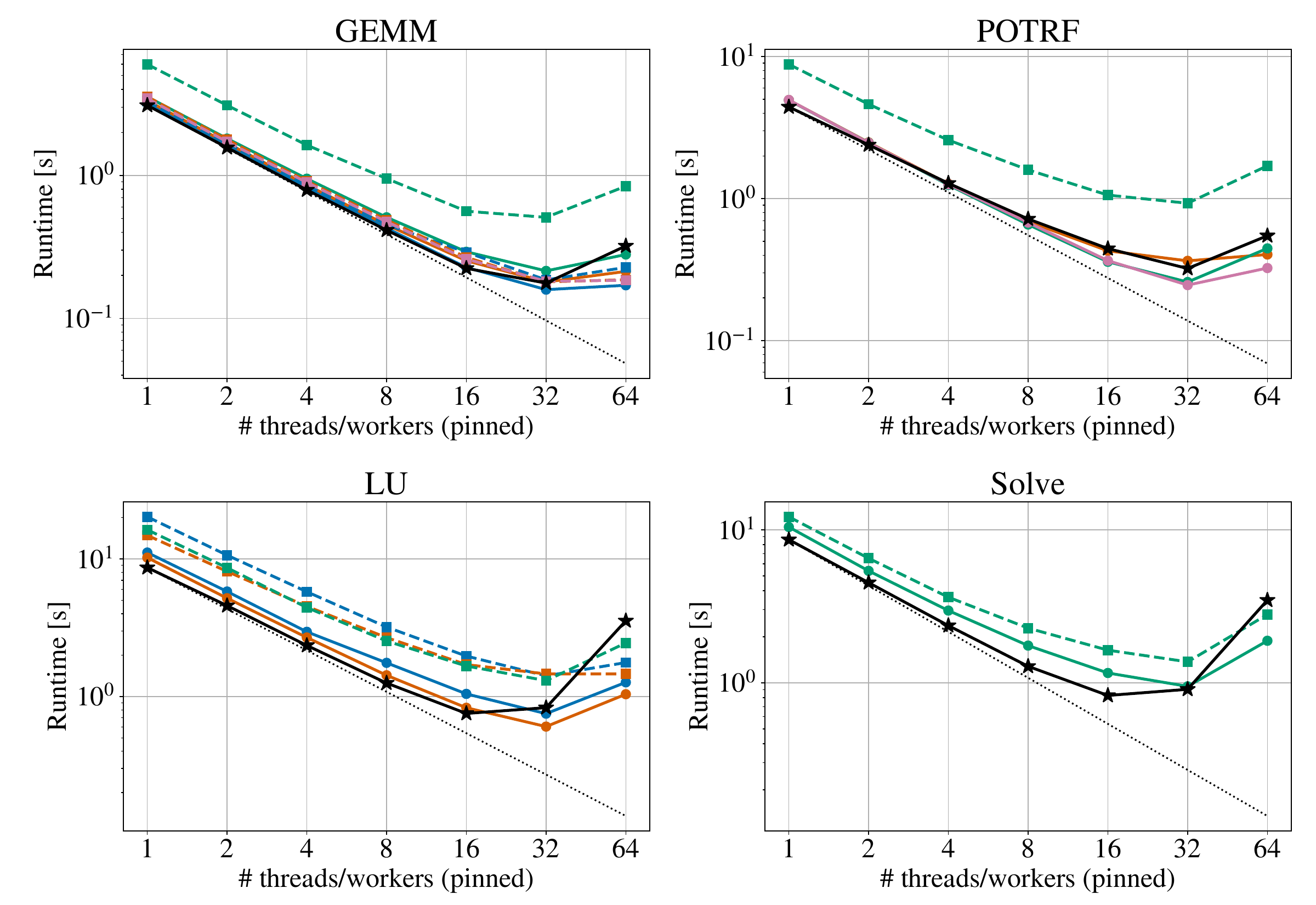}\\[2ex]
    \includegraphics[width=.95\textwidth]{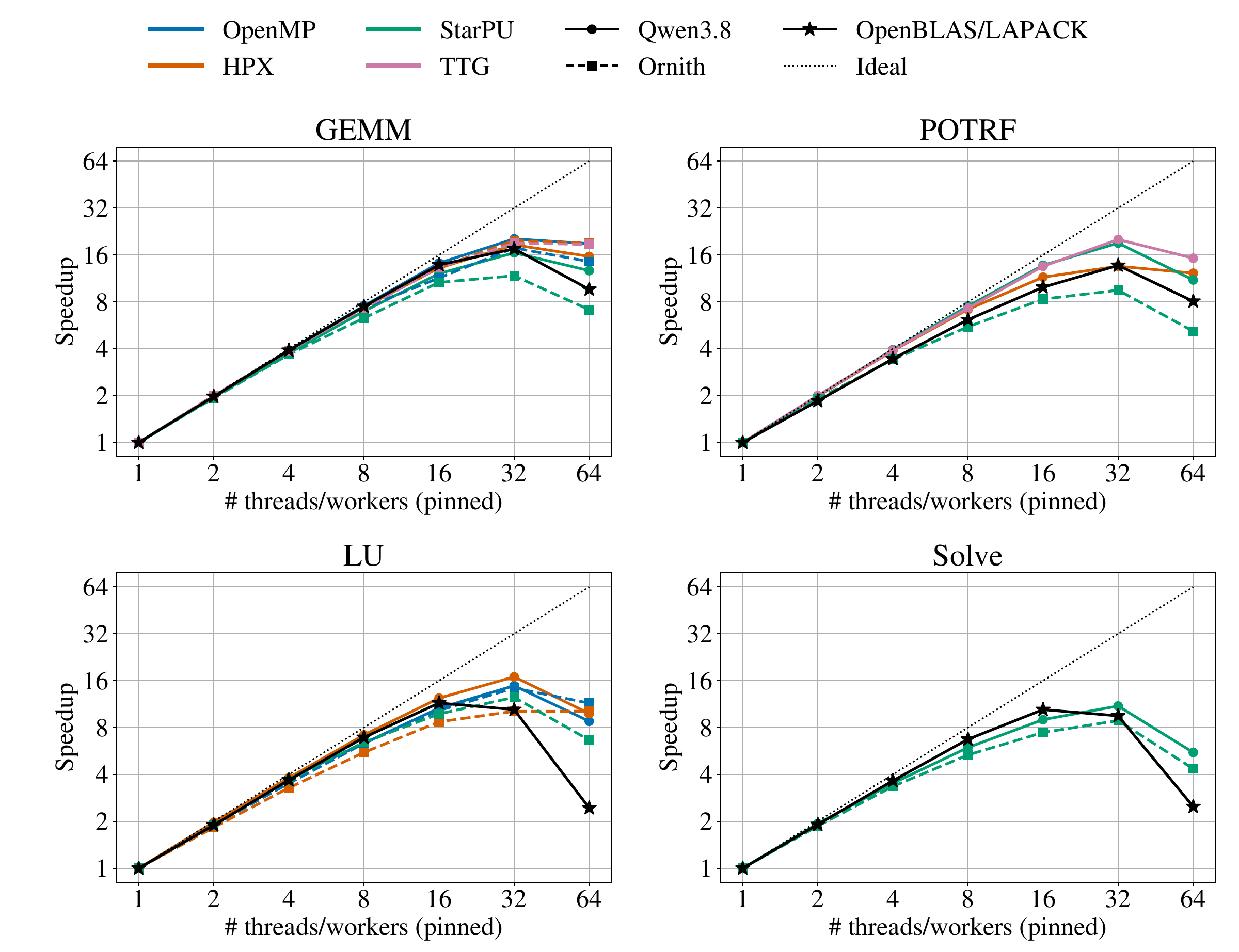}
    \caption{Absolute runtime (top) and self-relative speedup (bottom) of the 18 selected kernels and the OpenBLAS/LAPACK reference with $p$ threads pinned to the cores $0,\dots,p-1$. Colors denote the framework, line styles the model; dotted: ideal scaling of the fastest single-core run.}
    \label{fig:pinned}
\end{figure}

\paragraph{Pinned Runs of the Best Kernels.}
Without pinning, the threads are distributed over both sockets, which penalizes the OpenBLAS reference in particular.
We therefore re-measure the reference and the 18 correct parallel GEMM and factorization kernels whose runtime on 32 cores is at most twice that of the unpinned reference, all by qwen3.8:27b and ornith:35b, with $p$ threads pinned to the cores $0,\dots,p-1$ (\Cref{fig:pinned}).
With pinning, the reference achieves speedups of 17.5 for GEMM and 13.7 for Cholesky on 32 cores, while LU and Solve are fastest on 16 cores and drop to a speedup of 2.5 on both sockets.
The selected kernels reach speedups of 9 to 20 on 32 cores, and 10 of them are faster than the reference on 32 cores or at most 5\,\% slower.
The fastest generated kernels beat the best reference runtime for GEMM (0.159\,s vs.\ 0.176\,s, OpenMP), Cholesky (0.246\,s vs.\ 0.322\,s, \ac{ttg}), and LU (0.604\,s vs.\ 0.753\,s, HPX), all by qwen3.8:27b, but not for Solve (0.948\,s vs.\ 0.827\,s).
On 64 cores, the tiled task-based kernels lose less than OpenBLAS, e.g., the LU kernels retain speedups of 6.6 to 11.5.

Finally, we use the PCGQS~\cite{bantel_prompts} to evaluate the tested \acp{llm} over all problems and parallel frameworks.
qwen3.8:27b and ornith:35b achieve the highest scores with 0.67 and 0.64, while qwen3.6:35b reaches only 0.43, since its kernels without parallelism are scored as incorrect.
nemotron3:30b and nemotron-3-nano:4b score 0.08 and 0.00, respectively.
Regarding memory, only the HPX AXPY kernel by nemotron3:30b copies its input; all other correct kernels stay within 250\,MB of the peak \ac{rss} of their serial baseline.

\section{Conclusion}\label{sec:conclusion}

We evaluated \ac{llm} agents that write task-parallel C\texttt{++} for OpenMP Tasking, HPX, StarPU, and \ac{ttg}.
Compared to single one-shot prompts, the agentic workflow substantially changes the picture: capable models compile and test their code and thereby resolve most API errors on their own, which yields correct kernels for 63--86\% of all configurations without any human intervention.
The choice of model, however, dominates: models that do not build and test their code almost never succeed.
Among the task-parallel frameworks, OpenMP Tasking and HPX are handled most reliably, StarPU is close behind, and \ac{ttg}, the youngest and least documented framework, trails clearly.
Success thus tracks the familiarity of the API more than the difficulty of the parallel algorithm.

At the same time, correctness-based evaluation alone overstates the quality of the generated code.
Several ``parallel'' kernels contain no parallelism at all, and while the best kernels match the OpenBLAS reference, absolute runtime as well as speedup, at sufficiently large problem sizes, the speedup of the memory-bound kernels and of many StarPU and \ac{ttg} kernels remains low.
The causes are classical errors, i.e., serialized task graphs, unsuitable task granularity, and fork-join patterns, which the agents miss because they verify correctness but rarely measure scaling.

\subsection*{AI Use Disclosure}

Generative AI tools, including Claude~\cite{claude}, were employed to run and analyze the experiments, to create the figures, and to draft and edit the manuscript.
All AI-assisted outputs were carefully reviewed, verified, and edited by the authors to ensure factual accuracy, interpretive rigor, and scholarly integrity.
The final manuscript reflects the authors' original intellectual contributions and analytical work.
\bibliographystyle{ieeetr}
\bibliography{bib}

@misc{code_gen_1,
      title={Evaluating Large Language Models Trained on Code}, 
      author={Mark Chen and Jerry Tworek and others},
      year={2021},
      eprint={2107.03374},
      archivePrefix={arXiv},
      primaryClass={cs.LG},
      url={https://arxiv.org/abs/2107.03374}, 
}

@inbook{code_gen_3,
   title={Evaluating AI-Generated Code for C++, Fortran, Go, Java, Julia, Matlab, Python, R, and Rust},
   ISBN={9783031902000},
   ISSN={1611-3349},
   url={http://dx.doi.org/10.1007/978-3-031-90200-0_20},
   DOI={10.1007/978-3-031-90200-0_20},
   booktitle={Euro-Par 2024: Parallel Processing Workshops},
   publisher={Springer Nature Switzerland},
   author={Diehl, Patrick and Nader, Noujoud and Brandt, Steve and Kaiser, Hartmut},
   year={2025},
   pages={243–254} }

@inproceedings{parallel_kernels, series={ICPP-W 2023},
   title={Evaluation of OpenAI Codex for HPC Parallel Programming Models Kernel Generation},
   url={http://dx.doi.org/10.1145/3605731.3605886},
   DOI={10.1145/3605731.3605886},
   booktitle={Proceedings of the 52nd International Conference on Parallel Processing Workshops},
   publisher={ACM},
   author={Godoy, William and Valero-Lara, Pedro and Teranishi, Keita and Balaprakash, Prasanna and Vetter, Jeffrey},
   year={2023},
   month=aug, pages={136–144},
   collection={ICPP-W 2023} }

@inproceedings{HPC-Coder,
   title={HPC-Coder: Modeling Parallel Programs using Large Language Models},
   url={http://dx.doi.org/10.23919/ISC.2024.10528929},
   DOI={10.23919/isc.2024.10528929},
   booktitle={ISC High Performance 2024 Research Paper Proceedings (39th International Conference)},
   publisher={IEEE},
   author={Nichols, Daniel and Marathe, Aniruddha and Menon, Harshitha and Gamblin, Todd and Bhatele, Abhinav},
   year={2024},
   month=may, pages={1–12} }

@article{parallel_kernel_3,
  author       = {Godoy, William F. and Valero‐Lara, Pedro and Teranishi, Keita and Balaprakash, Prasanna and Vetter, Jeffrey S.},
  title        = {Large language model evaluation for high–performance computing software development},
  doi          = {10.1002/cpe.8269},
  url          = {https://www.osti.gov/biblio/2474767},
  journal      = {Concurrency and Computation. Practice and Experience},
  issn         = {ISSN 1532-0626},
  number       = {26},
  volume       = {36},
  place        = {United States},
  publisher    = {Wiley},
  year         = {2024},
  month        = {09}}

@inproceedings{Nichols_2024, 
   series={HPDC ’24},
   title={Can Large Language Models Write Parallel Code?},
   url={http://dx.doi.org/10.1145/3625549.3658689},
   DOI={10.1145/3625549.3658689},
   booktitle={Proceedings of the 33rd International Symposium on High-Performance Parallel and Distributed Computing},
   publisher={ACM},
   author={Nichols, Daniel and Davis, Joshua H. and Xie, Zhaojun and Rajaram, Arjun and Bhatele, Abhinav},
   year={2024},
   month=jun, pages={281–294},
   collection={HPDC ’24} }

@inproceedings{bantel_prompts,
author="Bantel, Linus
and Strack, Moritz
and Strack, Alexander
and Pfl{\"u}ger, Dirk",
editor="Diehl, Patrick
and Rampp, Markus
and Laure, Erwin
and Schulz, Martin",
title="From Prompts to Performance: Evaluating LLMs for Task-Based Parallel Code Generation",
booktitle="Asynchronous Many-Task Systems and Applications",
year="2026",
publisher="Springer Nature Switzerland",
address="Cham",
pages="15--26",
isbn="978-3-032-27676-6"
}

@article{hpx,
doi = {10.21105/joss.02352}, 
url = {https://doi.org/10.21105/joss.02352},
year = {2020}, publisher = {The Open Journal},
volume = {5}, number = {53}, pages = {2352},
author = {Kaiser, Hartmut and Diehl, Patrick et al.},
title = {HPX - The C++ Standard Library for Parallelism and Concurrency},
journal = {Journal of Open Source Software} }

@article{starpu,
author = {Augonnet, Cédric and Thibault, Samuel and Namyst, Raymond and Wacrenier, Pierre-André},
title = {StarPU: a unified platform for task scheduling on heterogeneous multicore architectures},
journal = {Concurrency and Computation: Practice and Experience},
volume = {23},
number = {2},
pages = {187-198},
doi = {https://doi.org/10.1002/cpe.1631},
url = {https://onlinelibrary.wiley.com/doi/abs/10.1002/cpe.1631},
eprint = {https://onlinelibrary.wiley.com/doi/pdf/10.1002/cpe.1631},
year = {2011}
}

@inproceedings{ttg,
  author={Bosilca, G. and Harrison, R.J. and Herault, T. and Javanmard, M.M. and Nookala, P. and Valeev, E.F.},
  booktitle={2020 IEEE/ACM Fifth International Workshop on Extreme Scale Programming Models and Middleware (ESPM2)}, 
  title={The Template Task Graph (TTG) - an emerging practical dataflow programming paradigm for scientific simulation at extreme scale}, 
  year={2020},
  volume={},
  number={},
  pages={1-7},
  doi={10.1109/ESPM251964.2020.00011}}

@article{parsec,
	title = {{DAGuE:} A Generic Distributed {DAG} Engine for High Performance Computing},
	booktitle = {Proceedings of the Workshops of the 25th IEEE International Symposium on Parallel and Distributed Processing (IPDPS 2011 Workshops)},
	year = {2011},
	month = {2011-00},
	pages = {1151-1158},
	publisher = {IEEE},
	address = {Anchorage, Alaska, USA},
	author = {George Bosilca and Aurelien Bouteiller and Anthony Danalis and Thomas Herault and Pierre Lemariner and Jack Dongarra}
}

@article{openmp_tasks,
  author={Ayguade, Eduard and Copty, Nawal and Duran, Alejandro and Hoeflinger, Jay and Lin, Yuan and Massaioli, Federico and Teruel, Xavier and Unnikrishnan, Priya and Zhang, Guansong},
  journal={IEEE Transactions on Parallel and Distributed Systems}, 
  title={The Design of OpenMP Tasks}, 
  year={2009},
  volume={20},
  number={3},
  pages={404-418},
  doi={10.1109/TPDS.2008.105}}

@article{tiled_la,
title = {A class of parallel tiled linear algebra algorithms for multicore architectures},
journal = {Parallel Computing},
volume = {35},
number = {1},
pages = {38-53},
year = {2009},
issn = {0167-8191},
doi = {https://doi.org/10.1016/j.parco.2008.10.002},
url = {https://www.sciencedirect.com/science/article/pii/S0167819108001117},
author = {Alfredo Buttari and Julien Langou and Jakub Kurzak and Jack Dongarra}
}

@inproceedings{react,
      title={ReAct: Synergizing Reasoning and Acting in Language Models}, 
      author={Shunyu Yao and Jeffrey Zhao and Dian Yu and Nan Du and Izhak Shafran and Karthik Narasimhan and Yuan Cao},
      year={2023},
      eprint={2210.03629},
      archivePrefix={arXiv},
      primaryClass={cs.CL},
      url={https://arxiv.org/abs/2210.03629}, 
}

@inproceedings{swebench,
      title={SWE-bench: Can Language Models Resolve Real-World GitHub Issues?}, 
      author={Carlos E. Jimenez and John Yang and Alexander Wettig and Shunyu Yao and Kexin Pei and Ofir Press and Karthik Narasimhan},
      year={2024},
      eprint={2310.06770},
      archivePrefix={arXiv},
      primaryClass={cs.CL},
      url={https://arxiv.org/abs/2310.06770}, 
}

@inproceedings{sweagent,
      title={SWE-agent: Agent-Computer Interfaces Enable Automated Software Engineering}, 
      author={John Yang and Carlos E. Jimenez and Alexander Wettig and Kilian Lieret and Shunyu Yao and Karthik Narasimhan and Ofir Press},
      year={2024},
      eprint={2405.15793},
      archivePrefix={arXiv},
      primaryClass={cs.SE},
      url={https://arxiv.org/abs/2405.15793}, 
}

@misc{qwen3,
      title={Qwen3 Technical Report}, 
      author={An Yang et al.},
      year={2025},
      eprint={2505.09388},
      archivePrefix={arXiv},
      primaryClass={cs.CL},
      url={https://arxiv.org/abs/2505.09388}, 
}

@misc{opencode,
  author       = {{OpenCode Contributors}},
  title        = {{OpenCode: The Open-Source AI Coding Agent}},
  year         = {2026},
  howpublished = {\url{https://opencode.ai/}},

  }

@misc{llamacpp,
	title = {llama.cpp},
	url = {https://github.com/ggml-org/llama.cpp},
	publisher = {ggml},
	author = {Gerganov, Georgi},
	month = sep,
	year = {2026},
}

@misc{claude,
      title={Claude},
      author={{Anthropic}},
      year={2026},
      howpublished={\url{https://www.anthropic.com/claude}}
}

@misc{bantel_julia,
      title={Generated, Parallel, Scalable? A Study of Agentic AI-Generated Julia Code on Supercomputers}, 
      author={Linus Bantel and Anna-Lena Roth and Jonas Posner and Dirk Pfl{\"u}ger},
      year={2026},
      eprint={2606.16534},
      archivePrefix={arXiv},
      primaryClass={cs.DC},
      url={https://arxiv.org/abs/2606.16534}, 
}

@phdthesis{daiss2026irregular,
  address = {Stuttgart},
  author = {Dai{\ss}, Gregor},
  doi = {10.18419/opus-18228},
  eventdate = {2026-02-25},
  language = {eng},
  school = {Universit{\"a}t Stuttgart},
  supervisor = {Pfl{\"u}ger, Dirk},
  supervisorgnd = {14255037X},
  title = {From irregular parallelism to portable GPU kernels : enabling efficient task-based GPU programming with HPX, Kokkos and CPPuddle to accelerate stellar mergers},
  type = {Dissertation},
  year = 2026
}

\end{document}